# A model to predict image formation in the three-dimensional field ion microscope

Benjamin Klaes[1], Rodrigue Larde[1], Fabien Delaroche[1], Stefan Parviainen[1], Nicolas Rolland[4], Shyam Katnagallu[2], Baptiste Gault[2,3], and François Vurpillot[1,2]

Corresponding Author: francois.vurpillot@univ-rouen.fr

1. Normandie Université, UNIROUEN, INSA Rouen, CNRS, Groupe de Physique des Matériaux, 76000 Rouen, France.
2. Department of Microstructure Physics and Alloy Design, Max-Planck-Institut für Eisenforschung GmbH, Max-Planck-Straße 1, 40237 Düsseldorf, Germany
3. Department of Materials, Royal School of Mines, Imperial College London, Prince Consort Road, London, SW7 2BP, United Kingdom
4. Laboratory of Organic Electronics, Department of Science and Technology (ITN), Campus Norrköping, Linköping University, SE-60174 Norrköping, Sweden

**Abstract**

This article presents a numerical model dedicated to the simulation of field ion microscopy (FIM). FIM was the first technique to image individual atoms on the surface of a material. By a careful control of the field evaporation of the atoms from the surface, the bulk of the material exposed, and, through a digitally processing a sequence of micrographs, a three-dimensional reconstruction can be achieved. 3DFIM is particularly suited to the direct observation of crystalline defects such as vacancies, interstitials, vacancy clusters, dislocations, and any combinations of theses defects that underpin the physical properties of materials. This makes 3DFIM extremely valuable for many material science and engineering applications, and further developing this technique is becoming crucial. The proposed model enables the simulation of imaging artefacts that are induced by non-regular field evaporation and by the impact of the perturbation of the electric field distribution of the distorted distribution of atoms close to defects. The model combines the meshless algorithm for field evaporation proposed by Rolland et al. (Robin-Rolland Model, or RRM) with fundamental aspects of the field ionization process of the gas image involved in FIM.

## 1. Introduction

Field ion microscopy (FIM) allows for direct imaging of the surface of a material with true atomic resolution. In a FIM experiment, a needle-shaped specimen is placed inside an ultra-high vacuum chamber, into which a low pressure (<$10^{-3}$Pa) of an imaging gas, such as neon or helium, is inserted. FIM relies on the ionization of these imaging gas atoms from prominent positions above the specimen's surface, exploiting the effect of an intense electric field. The electric field is produced from the application of high voltage (a few kV) to the specimen prepared as a sharply pointed needle and cryogenically cooled (T<100K). When the radius of curvature at the needle's apex is sufficiently small (i.e. below 100nm), the very strong DC electric field enables direct ionization of the image gas in the vicinity of the surface. The specimen is placed in front of an imaging screen. The divergent surface electric field drives away the ions from the surface, creating a highly magnified projection of the atomic arrangement at the specimen's surface onto the screen. Each bright spot in FIM images is the result of tens of thousands of imaging gas ions, projected onto the FIM screen. More details on the technique can be found in numerous textbooks [1–3] .

From the 1950s, and for the following two decades, FIM was established as the first true atomic-scale microscopy technique[4], achieving direct space imaging of the positions of individual atoms[5] with a sub angstrom precision[6]. The ability to localize precisely the 3-D coordinates of individual atoms in direct space is expected to find important applications in materials science. For instance, early atomic-scale investigations of structural defects, such as vacancies and dislocations in metals[7–10], were enabled by FIM. The reconstruction of 3-D distribution of single vacancies caused by ion irradiation in W by FIM was the first of its kind [11,12].

The most difficult aspect of FIM studies is to efficiently retrieve the information contained within the collected images. A FIM image must be carefully interpreted to obtain the positions of individual atoms in 3-D in the real space. Additionally, special care must be taken to avoid misinterpretations and artefacts of the images. Surface diffusion effects [13] or artefacts of field-evaporation[14] are often suspected to cause image distortions. More fundamentally FIM images are snapshots of the specimen's surface at given steps of field evaporation, and retrieving 3D information can be tedious. Successful attempts at reconstructing 3D data volumes from FIM images have already been reported [15,16] and automated procedures were developed for the reconstruction of accurate 3D, lattice-resolved atom maps. These procedures consist in the application image processing techniques to detect the positions of hundreds of atoms within each micrograph and to digitally track them across a sequence of FIM images, from the moment when they are first revealed at the surface of the specimen until their field evaporation. Using the latest generation of graphic workstations, images extracted from FIM movies can been used to produce high resolution 3D reconstructions of the whole apex. This breakthrough qualifies theoretically FIM as the most precise available tool for the full tomographic real space imaging of atoms in bulk materials, in volumes larger than 100x100x100 nm$^3$. Nevertheless, to achieve this ultimate goal, the spatial performances of the technique (such as spatial resolution, accuracy, noise, efficiency…) must also be qualified on a theoretical basis.

To this aim, a preliminary model developed to interpret 3D reconstructed images get from FIM was recently developed [17]. In this model, it was shown that atomic positions reconstructed from FIM in a pure metal case (W) could be experimentally observed with systematic displacements of a maximum of 0.05 nm due to the change in the direction of the highest electric field induced at the surface caused by the field evaporation of neighbouring atoms. This was done by comparing experimental results with high electric field molecular dynamics simulations. This finding further shows that the observed atomic displacements in FIM are due to the field redistribution rather than due to real atomic movements. This first work has provided more insights into the physics of image formation in FIM. In addition, this result shows the relevance to take into account the realistic field distribution at the surface of FIM sample, compared to a simple model assuming geometrical considerations as it was developed in the past [18,19]. We propose in this paper to extend and improve this model to enable to understand image FIM formation in more realistic and general cases adapted to the formation of the image in the three-dimensional field ion microscope. A comparison of experimental and simulated images is provided. Comparisons are provided on tungsten which is the standard material for FIM. The ultimate spatial resolution and detection efficiency were estimated based on the model and compared to experiments. The quantitative agreement of the model with experiment is demonstrated, showing the ability of this model to push the performances of the 3D FIM to its ultimate limits.

## 2. Methods

### 2.1 Experimental details of the 3DFIM and theoretical consideration

In a FIM experiment, the specimen of end radius $R$ is biased to a potential $V$ of several kilovolts. Due to lightning-rod effect, an intense electric field is produced at the apex. This electrostatic field is inversely proportional to radius of curvature at the specimen's apex as shown by the Eq.(1).

$$F = \frac{V}{k_F R} \quad (1)$$

Where $k_f$ is a geometric factor [20].

Under the effect of the intense electrostatic field, the gas atoms are polarized and then attracted towards the surface of the specimen. When an atom is close enough to the surface, it can be ionized and accelerated by the electric field towards the detector system. Image gas ionization is generally understood as a quantum process of electron tunnelling from the gas atom to the tip surface. Note that gas atom is assumed to be thermalized to the cryogenic sample temperature. This effect was experimentally evidenced by E.W. Müller in 1951 [21] and the theory was developed in the following years. An electron from the last occupied state inside the free gas atom sees the coulomb attractive force of the ion core. Electron occupies a quantified electronic energy level inside the atom defined by the ionization energy **I** (first ionization energy). The field *F* bends the energy curves and an energy barrier in the free space between the surface and the atom exists (Fig. 1).

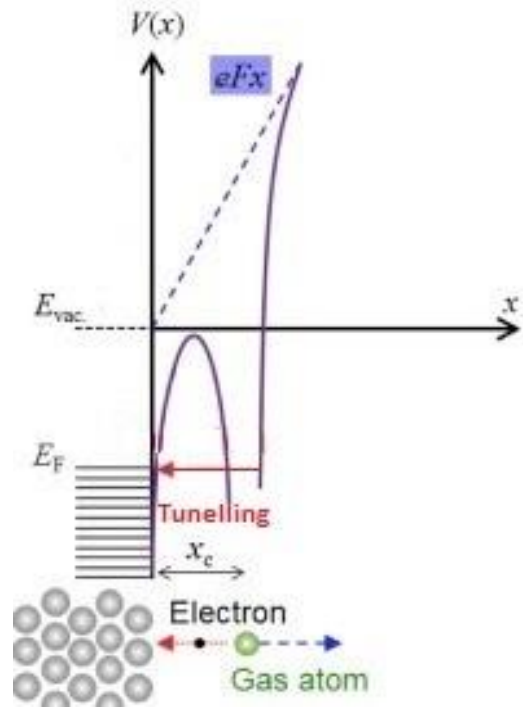

***Figure 1*** *: Field ionization by tunneling effect under an high electrical field. The electrical field (F) generated by the electric potential applied to the tip (V) deforms the potential barrier in which the electrons of the image gas atoms are trapped. At a critical distance xc, where gaz electron energy level is aligned with the metal Fermi level, tunneling probability of electron is maximal. If the distance between the gas atom and the surface of the tip is smalled no free electron state exists and tunnel effect is prohibited.*

Ionization is produced by electron tunneling across the energy barrier existing between the atom and the metal surface. The tunneling transmission is calculated using classical Jeffreys-Wentzel–Kramers–Brillouin (JWKB) method [22] . When the atom is close to the tip, the barrier height is further reduced by the effect of the electron's positive image and the negative image of the ion behind the conductor surface, and the width of the barrier is reduced by the presence of the Fermi level of the material of interest. The ionization probability is maximum when the atom is located at the distance $x_c$ named critical distance of ionization. This distance is the smallest distance for which ionization of the free atom in front of the material can occur.

Assuming the Fermi level $\phi$, if the atom is at the distance smaller than

$$x_c = \frac{I - \phi}{eE} \qquad (2)$$

Several refined models have been developed to give an approximate accurate analytical expression for the optimal tunnelling rate (at $x=x_c$). The most well-known is the equation provided by Gomer [23]:

$$T(F, x_c) \sim exp\left(-6.83 \times 10^9 \left(\frac{I - \phi}{F}\right) \times \left(1 - 7.6 \times 10^{-5} \frac{F^{1/2}}{I}\right)^{1/2}\right) \quad (3)$$

where I and $\phi$ in eV, and F in V/m

Each surface atom is seen as a protruding feature with a higher local curvature than the average curvature of the specimen, thereby enhancing the local electric field. These surface atoms are also places where the gas atoms can adsorb and further enhance the roughness of the surface, and consequently the local field [24,25].

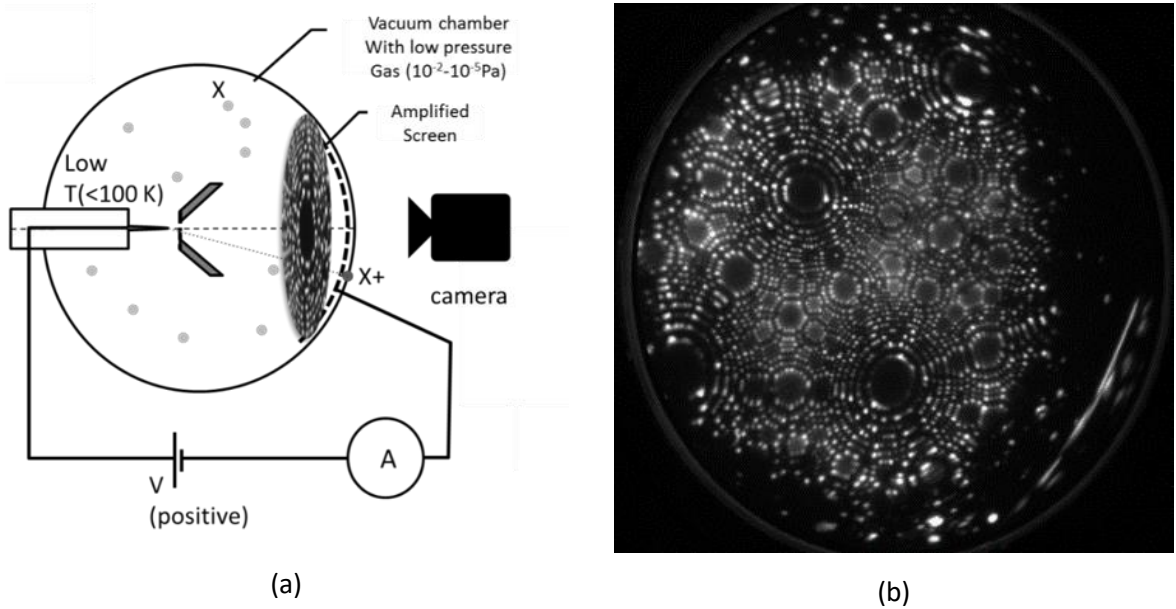

(a) (b)

*Figure 2 : Field ion microscopy experimental setup (a). Tip is placed in a vacuum chamber with a low pressure gas (He, Ne,…) and cooled at a temperature below 100K. Field ion micrograph (b). Each bright spot on the screen corresponds to an atom on the surface of the tip apex.*

Above the surface at the critical distance of field ionization, each surface atom is seen as a local site for ion gas emission. This beam of ions (a few thousand per second) is emitted, and accelerated by the electrostatic field. At the mesoscopic scale the field distribution is strongly diverging and a magnified image of the surface is produced on the imaging screen placed at 3–10 cm away from the specimen.

Note that this last expression is an approximation of the actual tunneling rate above the sample surface. For instance, it does not take into account the presence of an adsorbed layer on the surface that is experimentally of great influence [26]. The influence of the real electronic structure of alloys is also not taken into account. More refined theories are necessaries to fully address the contrast existing on the surface. For instance tunneling probabilities calculated using the Tersoff-Hamann approximation shows that the electronic structure greatly influences the tunneling rate [27]. Indeed, at the critical distance of ionization the tunneling rate is proportional to the surface local density of states, which may give additional enhancement of the contrast in the case of alloys. In this paper, we will neglect these enhancements, but further modifications of eq. 3 are possible to implement it in.

The imaging system is composed of a microchannel plate, to amplify the signal, and a phosphor screen, to convert the electron signal in photon signal. On the phosphor screen it is possible to observe a projection of the specimen's surface with a high magnification ($\sim 10^6$). As represented in the Fig.(2.b) each bright spot corresponds to a projection of a surface atom position at the surface of the tip. At low temperature, the size of spots is about 0.2 to 0.5 nm in diameter. The concentric rings are created by the intersection of the crystallographic planes on the material and the hemispherical surface of the tip apex.

In three dimensional FIM experiments, it is necessary to image the surface and field evaporate the specimens at the same time. The applied DC voltage is controlled to get slow field evaporation of the material (from $10^{-2}$ to $10^5$ atom/s in the general case). In some special cases, like with tungsten, the evaporation field is too high and, when this field is reached, atomic resolution on the images is lost because ionization of gas atoms is produced far from the tip surface. It is possible to resolve this problem using additional short nanosecond voltage pulses superimposed to the DC voltage [16]. The FIM contrast is then controlled by the DC voltage and the controlled field evaporation is triggered at or near to the top of voltage pulses. A regular evaporation rate is maintained during all the experimentation with a feedback control on the total (DC+pulsed) voltage. The appropriate ratio of the amplitude of the high-voltage pulses to the DC voltage is material dependent ($V_P/V_{DC}$ around 35% for tungsten).

*2.2 Image reconstruction*

Here, an algorithm modified from that of Dagan et al. [16] was used to retrieve the position of the atoms on the screen, and thereafter perform a reconstruction in the analysed volume. These sequential 2D hit positions extracted from the FIM images formed a space called "image space" that we refer to as the "detector space". Applying a reverse projection law from the coordinates of these positions allows to obtain a reconstruction of atomic positions.

First, the raw images are optimized by image processing. The micrograph in Fig.(3a) was obtained on a tungsten wire oriented, along the <011> direction, and prepared by electrochemical-polishing with a solution of 5% NaOH in water with a voltage in the range of 5–8V. FIM was performed at 40K with a pulse fraction of 35%, and He as an imaging gas, under a pressure of $10^{-3}$Pa. Several million of atoms were collected from the controlled field evaporation of the specimen. A region of interest is cropped and showed in Fig.(3b). Brightness and contrast are adjusted, a noise thresholding is applied using a representative intensity histogram of the set of equalized images. The best threshold is chosen as the minimum between the peak corresponding to the noise and the peak of the actual spot signals. A median filter is used to reduce the local noise level and images are smoothed with a mean and a Gaussian filter ($\sigma\sim 1$ pixel). The final result, following image processing, is presented in Fig.(3c).

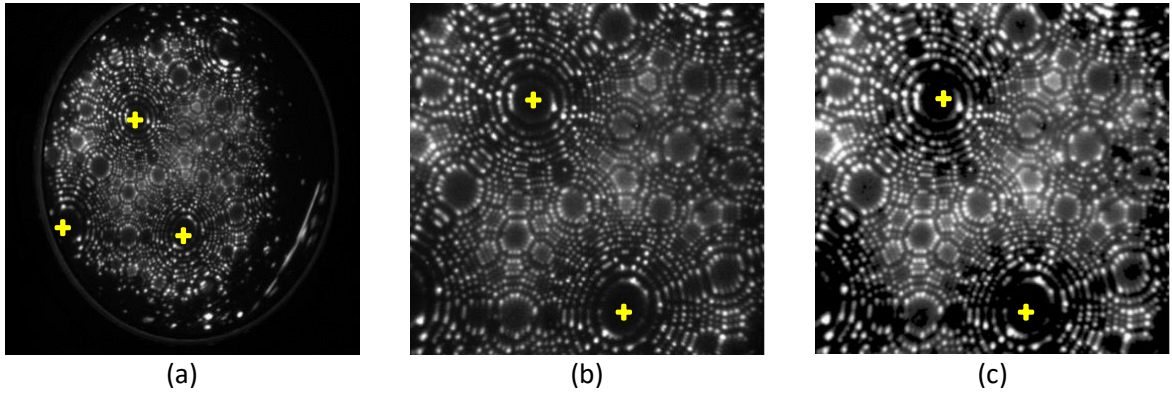

*Figure 3 : Raw image of tungsten record with CCD camera at 3.5 fps, (011) poles are identify by yellow crosses (a). Cropped image on a region of interest where the resolution is enough to apply the recognition algorithm (b). Final image after contrast improvement and noise reduction steps (c).*

Second, atom identification is performed by recognition of sudden changes in contrast between two successive images that correspond to where atoms were field evaporated (Fig 4a,b). Indeed, at a slow evaporation rate, the contrast between two successive images seldom varies, and the only source of significant variation is the removal of a limited number of atoms. When an atom is removed, the contrast locally changes from bright to dark at this precise location. The direct subtraction of image intensity map between two successive images reveals that field evaporation occurred at specific positions. The resulting differential image in Fig.(4c) presents bright contrasts at the location where atoms were field evaporated.

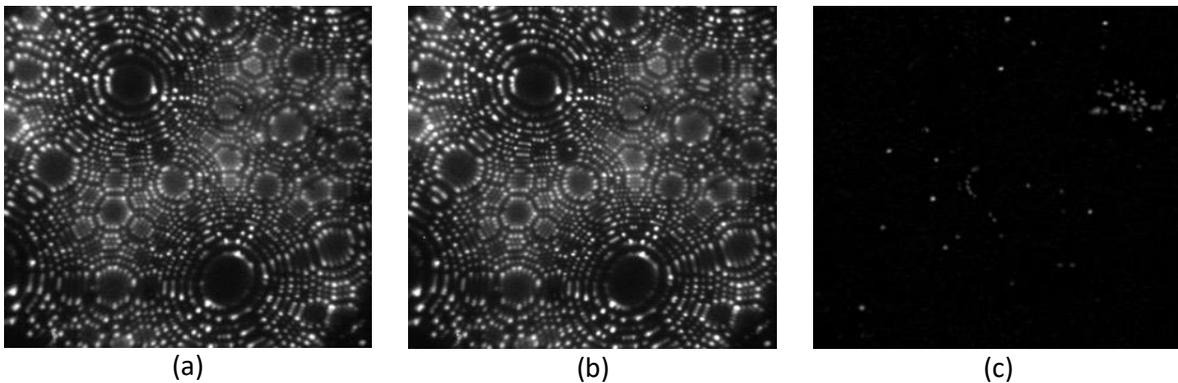

*Figure 4 : Results of the difference after thresholding (c) between an image take at time $t_1$ (a) and an image took at time $t_1 = t_0 + \Delta t$ (b). Each bright spot represent an evaporated atom in the time $\Delta t$.*

Then, the detection of local maxima on this resulting image is done in three steps. The first step uses an ordering (or sorting) filter that sorts pixels by grey level value in a given neighborhood around the considering pixel, i.e. similar to a median filter. More details can be found in ref. [28]. In this case we used an order filter of the 'maximum' type using a mask as showed in Fig.(5a). Pixels of the original image, showed in Fig.(5b), are scanned one after the other. For each of them, the algorithm looks in the neighboring pixels, contained in the mask having a value of 1, the pixel of the image which has the highest grey level. Then this value is put in place of the value of the analyzed pixel on the original picture. The results of this filter using the mask of Fig.(5a) applied on pixel of Fig.(5b) is illustrated in Fig.(5c). Note that the example illustrated by Fig.(5) is an idealised example.

The second step in the process is to compare, with a logical test, the original image pixels values and the new pixel values of the filtered image. For a given pixel, the grey level on the filtered image is below the grey level on the original image if this pixel corresponds to a local maximum. Pixels that correspond to this criterion are then marked, assigning it the value 1. All other pixels that do not respond to this criterion are marked with the value 0. The image finally obtained, illustrated by Fig. (5d), corresponds to the local maxima.

After these two steps, on field ion micrographs, each detect local maximum is composed of a group of several connected pixels. The final step is a clustering approach, where a label is given to each group of pixels and a centre-of-mass is calculated to define precisely the position of each atom.

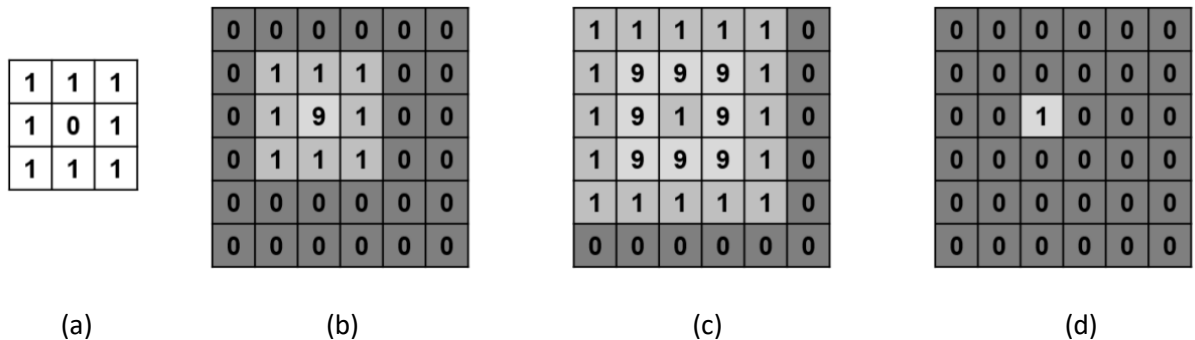

(a)　　　　　　　(b)　　　　　　　(c)　　　　　　　(d)

*Figure 5 : Main steps of the atom identification filter on a example of few pixels. Typical 3x3 pixels mask used to filtered images (a), pixels with value 0 in the mask are taken into account during the search for the maximum grey level. Image with a local maximum (b), value represents the grey of pixels. Filtered image result (c) after applying the order filter type maximum with 3x3 pixels mask. Final binary image obtained after a logical comparison between original images a filtered image (d), local maximum is located with value 1.*

After image processing, a count of the bright spots can give an estimated number of field evaporated atoms between two successive images, and an estimate of the evaporation rate. From this number, it is then possible to calculate the corresponding volume and so determine a depth increment between two successive images using conventional reconstruction procedures used in atom probe [29]. After this second step, a 3DFIM analysis consists on list of impact position on the image screen that formed a space called 'image space'.

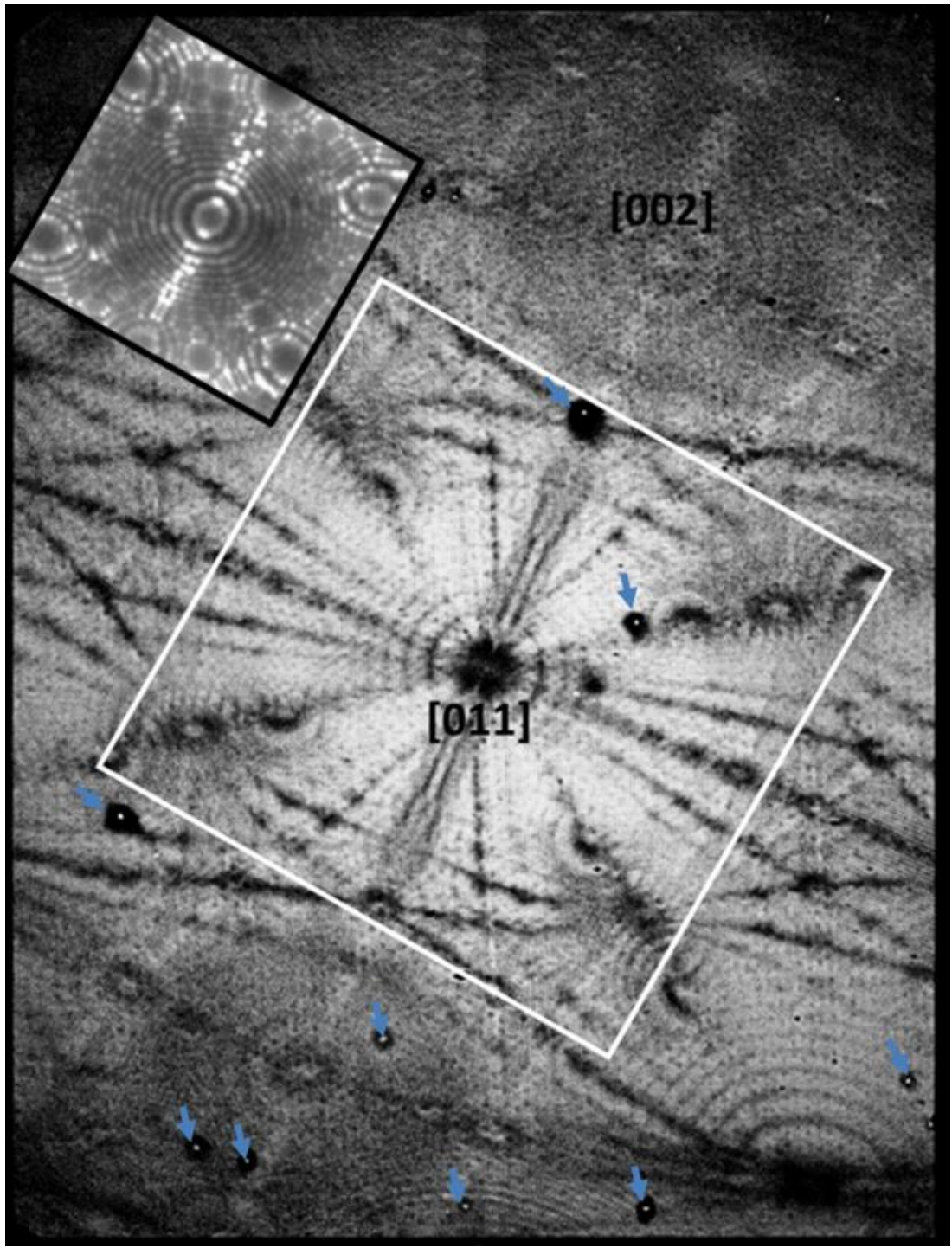

**Figure 6** : Stacking of the atoms positions on the detector (Detector hit map). Artefacts like concentric rings, dark and bright lines, induced by the crystallography-related faceting of the specimen, are clearly visible. Note that small dark spots (mark with arrows) are due to imperfection on the phosphor screen. Inset is shown the corresponding FIM image in the delimited white box

The transformation from the 'image space' into the specimen space consists in applying a reverse projection law. The projection law of the FIM is considered identical to the projection used in atom probe tomography. From the (X,Y) positions of impacts, (x,y,z) positions of atoms are reconstructed using the extended Bas protocol described elsewhere [29]. Note that the tip of the specimen is assumed to adopt a hemispherical shape of radius $R_i$. The evolution of the radius of curvature is determined assuming a constant shank angle evolution during field evaporation, or by using voltage V evolution to deduce R through Eq.(1) assuming constant field evaporation strength.

Fig.(7) shows a series of subset of the reconstructed volume from the dataset introduced in Fig.(6). The complete volume was $55 \times 55 \times 65\ nm^3$ containing around five million atoms. Each subset shows specific families of atomic planes and, along specific directions, atomic columns. After reconstruction, and adjusting the reconstruction parameters (using a constant field factor, an initial radius of curvature $R_0$ estimated from FIM images, the average atomic volume of tungsten atom of 1.57x10$^{-2}$nm$^2$, and global detection efficiency of 80%.) , the average interplanar distance between (011) atomic planes in the depth direction is $d_{011} = 2.24 \pm 0.05$ Å (close to the theoretical distance of tungsten 2.23 Å). Post reconstruction deformations exists but are small considering the simplicity the reconstruction algorithm. For instance, angle between the two 011 direction represented on the Fig.(7) is 84.5°, whereas it should theoretically be 90°. Distortions inherent to issues with reconstruction have been discussed in several papers [30,31].

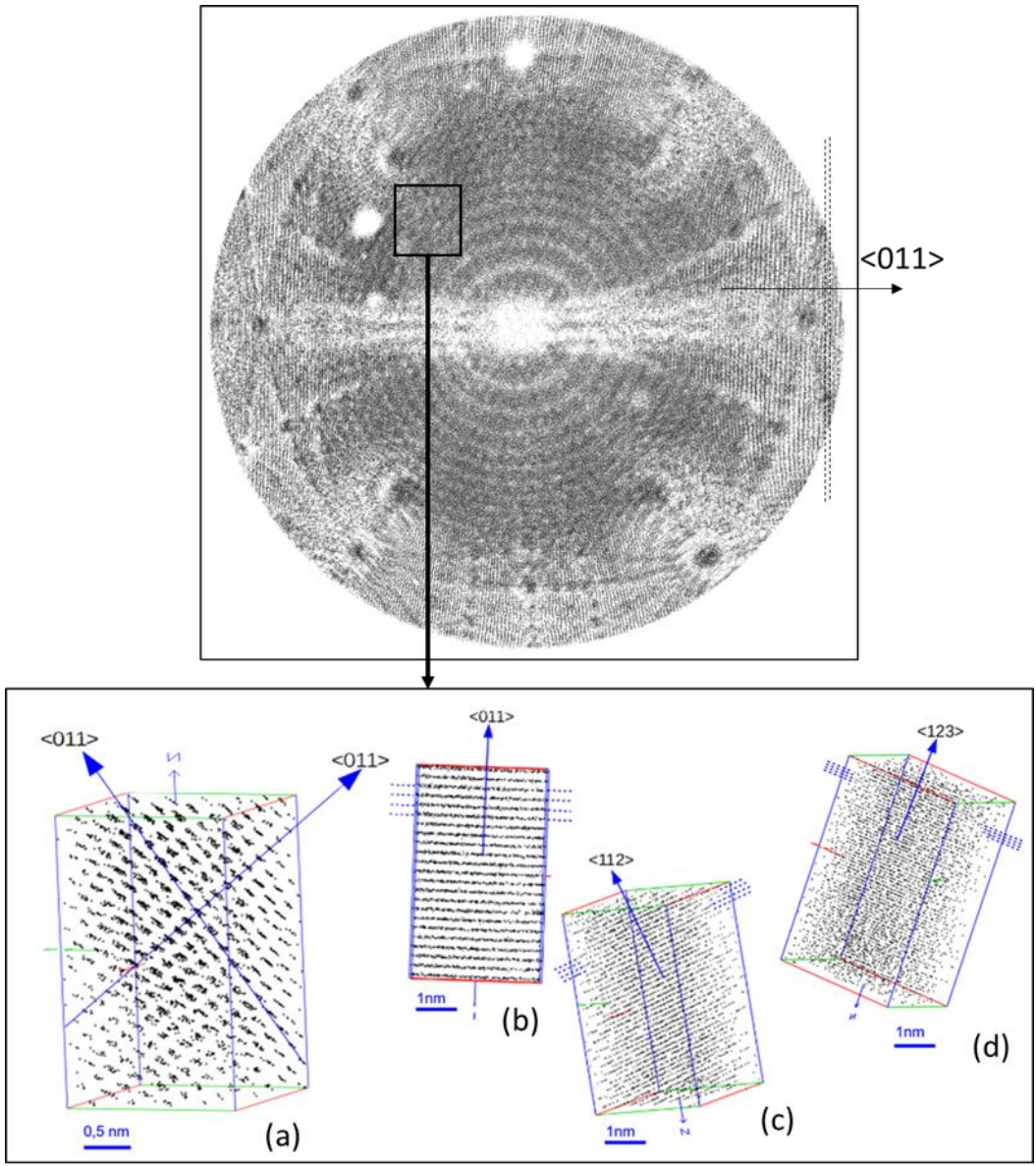

***Figure 7*** : *Small volume of $3 \times 3 \times 5\ nm^3$ selected in the complete reconstructed volume (top image) in different orientations. Several families of planes (b)(c)(d) and atoms column (a) are visible in some specific orientations of this volume. (011) planes are also reconstructed laterally on some regions of the whole reconstructed volume (e).*

### 2.3 FIM modelling

Modelling FIM images requires three steps. First, the atom-by-atom sequence of field evaporation from a virtual specimen must be calculated. This step is performed using the Robin-Rolland algorithm [32] developed for the interpretation of atom probe tomography imaging processes. Second, the tunneling probability of the image gas is calculated at different stages of the tip's field evaporation. Last, the projection of ions from the specimen's surface onto a plane simulating the projection on the screen or detector is evaluated. This last step requires an accurate calculation of the electric field distribution in the open space between the specimen and that plane.

### 2.3.1 Robin-Rolland field evaporation Model

The Robin-Rolland algorithm enables the dynamic calculation of the electric charge distribution at the surface of a virtual FIM specimen, assuming the charge being localised at the positions of surface atoms. In the present approach, the simulation space is only defined by the three-dimensional coordinates of each atom of the emitter and no additional support points (i.e. grid) are introduced, neither inside the virtual specimen nor the surrounding space. Classical electrostatic theory describes the perfect conductor object to a given voltage in vacuum as an equipotential surface enclosing a zero-field volume. This equilibrium is instantaneously reached through the redistribution of the free electrons from within the conductor. As a result, the volume charge density is strictly zero, while there is a non-zero surface charge density $\sigma(\vec{r})$, distributed all over the surface S of the virtual specimen.

The above calculations of the charge distribution are based on the assumption of a smooth surface of a conductive material considered as a continuous medium. As a result, the charged surface is considered as an infinitely thin layer, and classical electrostatics can be applied. In a material, the charge is found to concentrate on the atoms at the surface. These surface atoms may be considered as partial ions [33–36], with the most protruding atoms building the strongest charge, and therefore being the locations of the highest external field.

The elementary surface occupied by each atom is considered as a constant $S_{at}$. By applying a constant electrostatic potential, only the atoms (i=1,N) of the extreme surface will be subject to a non-zero surface charge density $\sigma_i$. The charge $q_i$ is given by the product $S_{at} \times \sigma_i$. To calculate this surface charge, the discretized form of the Robin equation is used :

$$\frac{q_i}{S_{at}} = \frac{1}{2\pi} \sum_{k=1, k \neq i}^{N} q_k \frac{\vec{n}_i \cdot \vec{r}_{ik}}{\|\vec{r}_{ik}\|^3} \tag{5}$$

Here $\vec{n}_i$ denotes the outward normal to the surface at the position of the atom *i*, and $\vec{r}_{ik} = \vec{r}_i - \vec{r}_k$.

For the purpose of finding the charge distribution over the field emitter's surface, one can now construct a sequence $q_{i,n}$ for each atom *i* . $\frac{q_{i,n+1}}{S_{at}} = \frac{1}{2\pi} \sum_{k=1, k \neq i}^{N} q_{k,n} \frac{\vec{n}_i \cdot \vec{r}_{ik}}{\|\vec{r}_{ik}\|^3}$ This sequence tends toward the charge distribution at equilibrium, at least for a convex surface. In the convergence process, the atomic surface is chosen to ensure that the total charge of the system remains constant. Once the convergence is reached, the charges are rescaled to give the correct applied voltage $V_{DC}$ using Coulomb's law.

This last equation represents a system of N coupled, linear equations which can be solved numerically. The brute force resolution of this system of equations exhibits a $O(N^2)$ complexity, N being the number of surface atoms. To avoid such a computational effort, equation (6) was optimized using a Barnes-Hut algorithm [37] approximating the long range interactions. In this approach an octree structure with the position of the N surface atoms was described reducing the complexity to ~ $O(NlogN)$. More details about the procedure is described in details in [32].

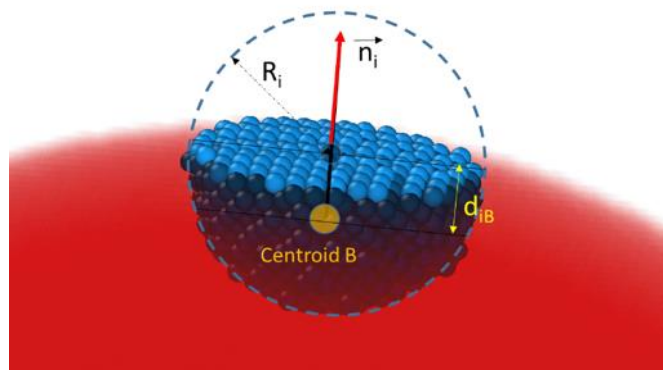

*Figure 8* : *Method used to estimate the surface normal to the tip composed of atoms. The centroid B of atoms situated in a small sphere of radius $R_i$ centred on atom $A_i$ is calculated. Surface normal vector $\vec{n}_i$ is the unit vector of direction $\overrightarrow{BA_i}$. $\|\overrightarrow{BA_i}\| = d_{iB}$ enables to discriminate surface and bulk atoms.*

We modified slightly the model to describe any arrangement of atoms inside the specimen's tip. Identification of surface atoms from arbitrary structures, as well as definitions of normal directions for these N surface atoms are required in Eq.(5). Surface atoms are atoms surrounded by an asymmetric distribution of neighbors. To quantify this asymmetry, a simple method derived from work of Boll et al. was developed. A sphere of radius $R_i$ is placed on all atoms of the simulated volume (Fig.(8)). The position B of the centroid calculated on all atoms inside the sphere, enables the calculation of the vector $\vec{u}_{Bi}$. Atoms for which $d_{iB}$ the length of $\vec{u}_{Bi}$, is higher than to a user-defined criterion $\delta_B$ are defined as surface atoms. The direction of $\vec{u}_{Bi}$ gives directly the normal direction to the surface. The choice of $R_i$ will be a compromise between a smooth definition of the normal direction and the computation time. Indeed, the higher is $R_i$, the larger is the number of atoms inside the sphere, and the more accurate is the estimation of the local normal. $R_i$ is chosen to maintain approx. 50–100 atoms inside the sphere. The user criterion $\delta_B$ is an ad-hoc parameter chosen to keep a monolayer of atoms at the tip surface. In this case, $\delta_B$ is fixed to 0.2 nm.

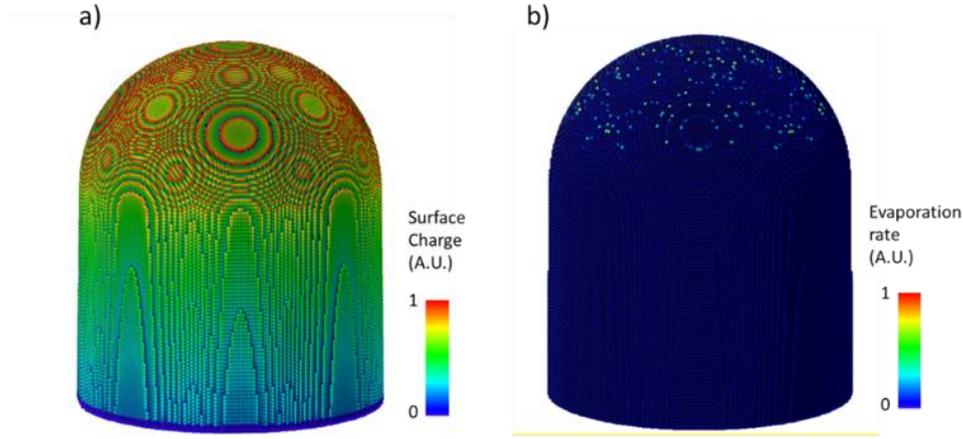

*Figure 9* : *Surface charge (a) on atom represented by a color scale. (b) Evaporation probability or rate calculated from eq. 9.*

Once the surface atoms are determined, an optimized version of Eq.(6) is iterated. The rapid convergence of the RRM, starting from a uniform value of charge, enables the calculation of the charge distribution across the virtual specimen's surface in less than a quite few iterations (convergence criteria <$10^{-10}$ in less than 100 iterations). We may note that error decreases exponentially with iterations. The external field is directly derived from the charge distribution at any place using Coulomb's equation. Once again, the octree structure is used to calculate this equation, ensuring an acceptable computation time. The electric field just above the surface is calculated directly from the surface charge provided by the model.

The field evaporation order is evaluated using the strength of the electric field (or charge since it is proportional) over all surface atoms, and using the evaporation field of each atom. We applied the procedure proposed by Vurpillot et al. [38]. The rate of field evaporation K($s^{-1}$) of a given surface atom $A_i$ is known to follow classical Boltzmann's equation

$$K_i(s^{-1}) \sim \nu_0 exp\left(-\frac{Q(F_i)}{k_B T}\right) \qquad (6)$$

$F_i$ being the surface electric field applied to the atom, and $Q(F_i)$ a field dependent energy barrier. In the simplest approach, on each atom $A_i$, it can be defined a unique expression for $Q(F_i)$ that follows a monotonous relationship with $F_i$ [2].

$$Q(F_i) \sim C_{Ai} \times \left(1 - \frac{F_i}{F_{Ai}}\right) \qquad (7)$$

Assuming $C_{Ai}$ ~1 eV an energetic parameter proportional to the sublimation energy for most of the elements, the evaporation rate is strongly dependent of the ratio $\frac{F_i}{F_{Ai}}$. $F_{Ai}$ is a parameter that depends on the elemental nature of elements, and of the bounding interaction energies with neighbors. The rate of evaporation $K_i$ can be calculated for all surface atoms, and the choice of the atom to be field evaporated can be taken using this table of rate. A Monte Carlo algorithm can be useful, as used in ref. [39], but at T<80 K, $C_{Ai}/k_B T$~150, which means the probability to be field evaporated is sufficiently high only for the few atoms with the highest ratio $\frac{F_i}{F_{Ai}}$ (Fig.(9b)). The low temperature used in FIM makes the field evaporation process strongly field dependent and quasi deterministic.

An estimate of $F_{Ai}$ is required in this algorithm. We generally assume a constant value for each element, that is calculated on the basis of the fields estimated for the pure elements [1]. Some approaches have been proposed to express an accurate expression that depends of the local environment of each atoms to calculate a local critical electric field to induce field evaporation [40–42], and this is expected to be integrated in future implementation of the model. When an atom is selected for field evaporation, it is simply removed from the surface, and a new list of surface atoms is generated. The surface charge of the evaporated, ionized atoms gets back to the total charge of the virtual specimen, and the distribution of surface charges is recalculated. The convergence of the algorithm is then extremely fast, since the charge of most of the atoms are not strongly modified by the removal of a single partial charge, i.e. 5-10 iterations are generally sufficient to re-calculate the charge distribution at the tip surface.

*2.3.2 Electron tunneling and ionization rate*

Once surface electric field *F* is known, it is straightforward to evaluate the electron tunnelling rate above each surface atom and then to determine the ionization rate. The critical distance of ionization and the tunnelling rate are calculated for all surface atoms through eq. (3) and (4). Obviously, we neglect tunneling that can occur farther from the surface, due to the rapid decrease of the field with the distance to the tip. Eq. 3 provide an approximation of the tunneling rate. If we consider the first ionization energy of He (I=24.5 eV), and a work function of 4 eV (usual value for metal surface), within a field range ±30% of the best image field (BIF=44 V/nm), eq.4 can be simplified to $T(f = F/BIF) \sim exp\left(-\frac{A}{f}\right)$, with A~14.82 (A is a field varying parameter, but variation with field is extremely reduced in the field range)..

In practice this semi-constant A can also be adjusted to provide better visual agreement with experiment (in the range 10–20, in this paper A was adjusted to 12) since the theoretical transparency also neglects a lot of second order parameters (gas adsorption layer, local density of state, etc…) .

*2.2.3 Projected image in FIM, 3D FIM and APT*

Atoms from the imaging gas are generally thermalized to the specimen's base temperature. As a result, we may consider that the initial velocity of the produced ion is almost zero (Ec~$k_B$T~2-10 meV in FIM). The position situated at the critical distance of ionization $x_c$, along the normal to the surface is chosen as the departure position of the center of the ion beam generated by field ionization. From this place, the trajectory of the ion is computed. The ion trajectory toward the detector is computed from the Newton motion equation as originally proposed by Vurpillot et al. [43], using a Runge Kutta numerical scheme. Although the calculation method makes it possible to compute the trajectory over any distance, we chose to calculate the trajectory up to 1 micron, and proceed further to conventional extrapolation at large distance up to a virtual planar detector.

Between the field evaporation of 2 atoms, the still surface of the virtual specimen is imaged onto the FIM screen. Each surface atom is the center of a beam of image gas ions that is projected by the electric field onto the screen. We can produce a map of this projection by plotting the FIM atom position map (or FIM hit map, Fig. 10a). Using these positions, a virtual FIM image is generated. Each atom location on the screen is used as a seed for a Gaussian distribution to mimic the distribution of ion intensity on the screen produced by a single atom at the tip surface. A FIM image is produced by the superimposition of thousands of light spot on the screen produced by the local ionization above surface atoms of the sample. Knowing the position of each ion beam on the detector, the tunneling rate must be evaluated for every ionization site. Assuming intensity being proportional to the tunneling rate, it is therefore trivial to generate a simulated FIM image. On the screen, we assume that each atomic position produces an intensity lateral distribution that follows a Gaussian shape. The standard deviation of this Gaussian is fitted to the experimental resolution of FIM image that depends on the specimen's temperature and image gas and varies between 0.2 to 0.5 nm [6]. The peak maximum of the Gaussian function is adjusted to the relative tunneling rate above each atoms. By evaporating the virtual specimen atom-by-atom, and computing the ion trajectories of every field evaporated atom (Fig. 10b ), a cumulative hitmap similar to those obtained by APT, as well as a 3D reconstruction of the evaporated volume can be calculated. At the same time, it is possible to store the FIM atom position of each field evaporated atom just before removing it from the surface (Fig. 10b). This position defines exactly the ideal 3DFIM atom position. By cumulating these positions for millions of atoms, an ideal 3DFIM hitmap (free from any detection algorithm artefacts) can be generated.

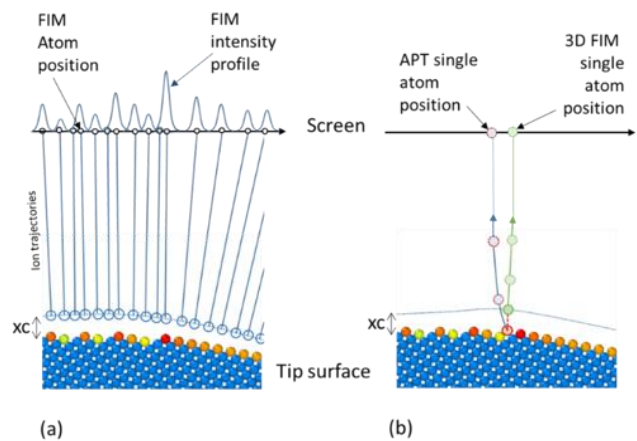

*Figure 10 (a) Schematic drawing of the calculation of FIM spot positions and FIM intensity map (a) and the calculation of the single position in APT and 3D FIM (b). In (a) the image is formed between evaporation of 2 atoms over the whole surface of the tip. Each surface atom location is projected to a ideal FIM hitmap. These positions are used to compute a virtual FIM image, using an intensity distribution for each FIM atom position.*

An example of the FIM hitmap produced by the computation of ion trajectories from the critical distance above each of the surface atoms is presented in Fig.(11a), in the case of a face centered cubic structure tip, with an embedded precipitate having a diamond structure and a relative evaporation field 20% higher than the matrix. The radius of the virtual specimen is 25 nm. Using these positions, and the electrostatic field distribution above the surface, a FIM image is generated (Fig. 11b). FIM hitmap (Fig 11a) and FIM image (Fig11 b) can be generated after the virtual field evaporation of thousands of atoms from the specimen. These images can be compared to the APT produced by field evaporating a thin layer of atoms, i.e. the equivalent of a hitmap collected in APT (Fig.11c). Because the field ionization occurs farther from the tip's surface compared to the field evaporation, the image gas ions and ions originating from field evaporation of the surface atoms do not follow the same trajectories. The typical artefacts that affect the spatial resolution in APT are hence visually less pronounced in FIM. These artefacts were quantified in a previous article [39].

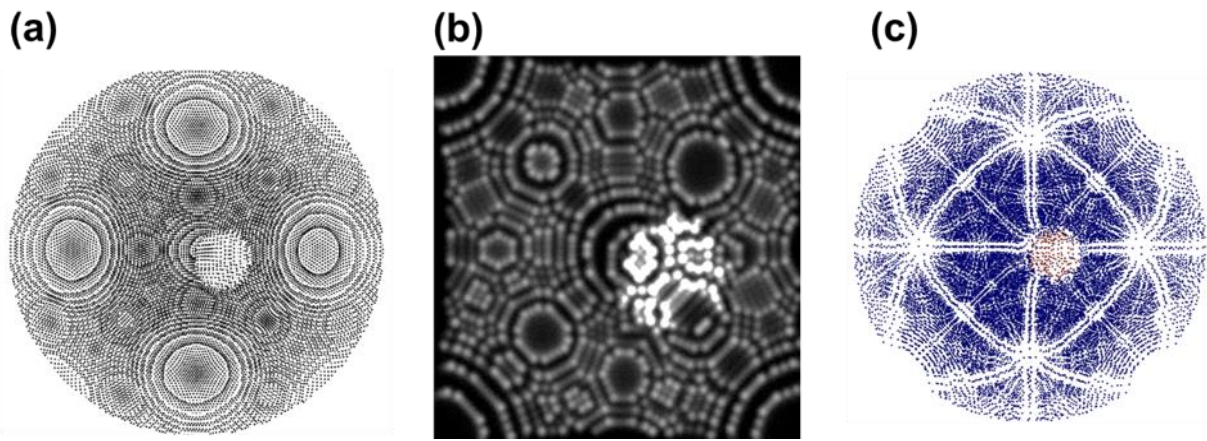

*Figure 11 (a) Ion hitmap of the positions of surface atoms in the case of a FCC cristal with a spherical diamond structure inclusion (field evaporation 20% higher than the matrix atoms) close to the tip centre, assuming a critical distance of field ionization given eq. (3) (no difference in electron work function between phases were assumed) (b) simulated FIM image, assuming the tunnelling probability and the critical distance of ionization. Difference in contrast between phases is uniquely linked to the calculated surface field (c) simulated APT hitmap for the evaporation of 3 atomic monolayers on the detectors. Trajectory aberrations of field evaporated atoms are strongly influenced by the charge distribution around evaporated atoms in the first steps of flight (generating zone lines contrast)*

### 3. Application to a pure metal case (tungsten)

#### 3.1 3D FIM simulation calculated on a [011] oriented tungsten tip

To assess the performances of the modelling tool in the prediction of image distortions in experimental FIM images, a BCC-tungsten virtual specimen was defined, with a diameter of 26 nm, aligned along the

<011> crystallographic direction. 500000 atoms were virtually field evaporated to produce a virtual atom probe dataset (APT reconstruction and hitmap), an ideal 3DFIM hitmap, and every 50 field evaporated atoms, virtual FIM images were generated (10000 images were calculated). Note that to reduce the computation time, a region of interest was selected on the detector, equivalent to a field-of-view of approx. 50° at the specimen. Results of these computations are presented in Fig.(12).

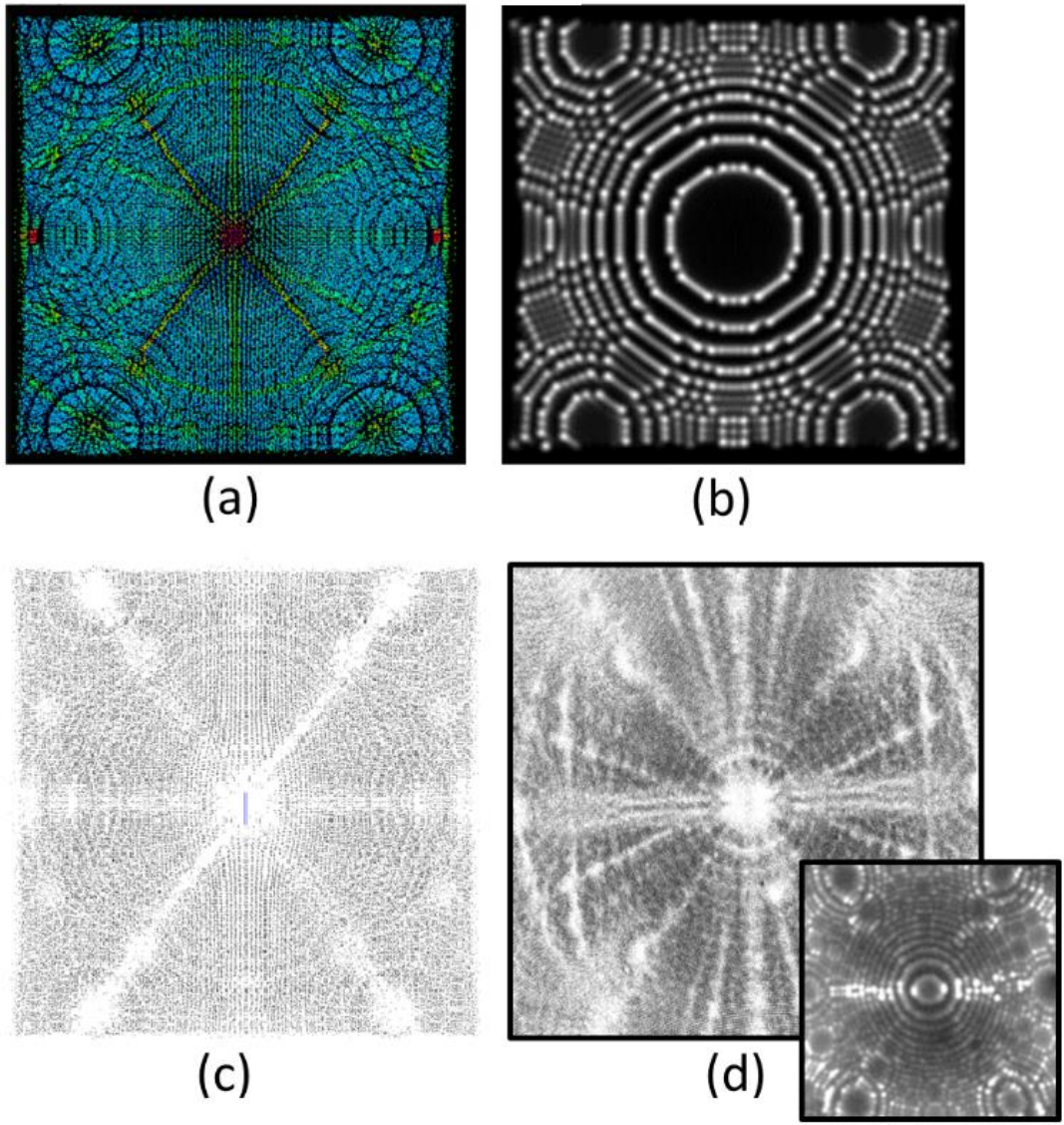

*Figure 12* FIM simulation of a tungsten tip, 26 nm in radius at the apex. 500000 atoms were field evaporated one by one following the proposed model (a) ideal 3DFIM hitmap on the screen, collecting 50° of field of view of from ions projected from the initial position above surface atoms just before field evaporation. Ion starting position is at $x_c$ from the surface, along the surface normal vector. Colormap indicated field strength at the moment of the field evaporation (Red=high field, Blue=low field). Note the presence of impact density variation induced by the dynamic local field distribution at $x_c$. (b)

*simulated FIM image at a given step of the tip field evaporation, assuming the tunnelling probability and the critical distance of ionization. (c) Hitmap produced by the detection of the initials positions of atoms obtained from the algorithm presented in section 2.2. The algorithm was applied to the sequence of simulated FIM images. Note the presence of additional depleted area on major zone lines and pole centres (d) comparison to the detection of atoms in a sequence of 10000 images produced following the evaporation of several millions of atoms on a [011] oriented tungsten tip (inset : Corresponding FIM image). Artefacts are strongly similar to the modelling approach.*

As observed in Fig.(12a), even if a single snapshot of ion impact (produced by ionization of ions above surface atoms) is rather homogeneous (Fig.(11a)). The accumulation of impacts generates significant distortions. Concentric density contrasts around the main low index poles are readily visible. Note that a perfect image should be almost homogeneous considering that only about ten (011) atomic planes were field evaporated and superimposed on the virtual detector. This kind of contrast is similar to the ring contrasts observed in field desorption images introduced by Waugh et al. in ref. [44]. They suggested that this ring contrast was the results of a periodic variation in the magnitude of "aiming errors", or of deflections of the metal ions, as the evaporating planes reach particular diameters. The rings are always of approximately the same size as the rings in the field-ion image, and obey the same law as the size of the rings observed by FIM, i.e. the diameter of the $n^{th}$ ring from the centre varies as $n^{1/2}$. When the last terrace at the centre of a pole is smaller than a critical size, the plane will evaporate abruptly, giving rise to a flat surface oven a significant size. The alteration in the magnitude of the local field at a plane edge is accompanied by a small change in the direction of the local field [17]. This feature, is smoothed at the critical distance of ionization but not sufficiently, giving rise to an average variation of the magnification at defined and regular places around the pole centre. A clear visualisation of this process can be observed in both simulation and experiment when looking at the 3DFIM contrast in the direction of the analysis. This is done by inserting an orthogonal slice in the 3D volume obtained by stacking-up a sequence of field evaporated images [45]. In Fig.(13a), an orthoslice was placed close to the pole centre in the simulated field evaporation image. Two (011) atomic planes were field evaporated during the sequence. The same procedure was applied to an experimental dataset (Fig.(13b)). In these figures, a bright filament along the depth axis is produced by the terrace field evaporation. The last atoms of the terrace at the pole centre evaporate steadily, and it is observed in both simulation and experiment a slight shrinkage of the concentric ring around the pole. This shrinkage affects several successive terraces. This feature appears regularly giving rise to a periodic contraction of the magnification around the pole.

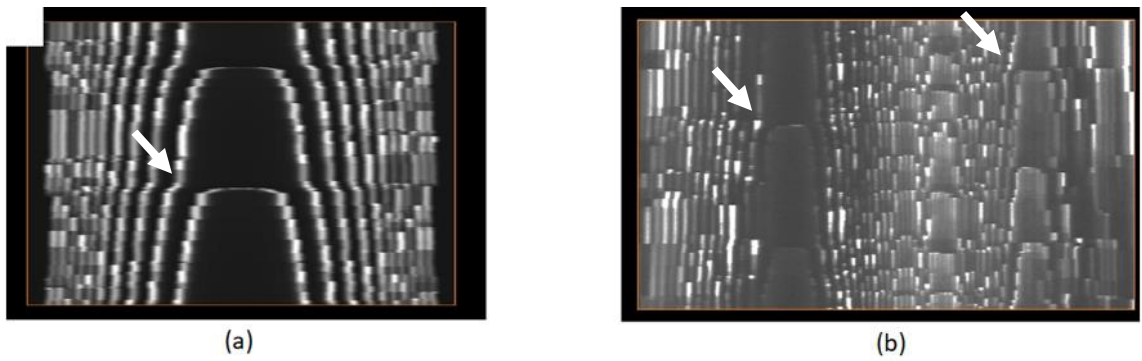

***Figure 13*** *Simulated (a) and experimental (b) FIM contrasts in slices parallel to the evaporation direction centred on the (011) pole. Two (011) atomic planes were fully evaporated. Shrinkage effects on terraces are visible on both simulated and experimental images (white arrows)*

In addition to these contrasts, zone lines and pole centres show variations in the density of impacts. When the last atoms from these regions are field evaporated, there are locally only few atoms on the respective atomic rows or planes. The local roughness varies strongly when a single atom is removed, giving rise to small deviations of the electric field lines as it was demonstrated in a previous study [17]. The sequence of field evaporation is however regular and periodic, so that the average deviation remains systematic, which changes the local magnification at these particular positions. Note that the impact of these deviations is almost invisible on still FIM images. Only an image-by-image analysis of FIM is necessary to highlight this phenomenon which is why it has seldom been studied up to now.

The calculation of a sequence of FIM images at increasing depths enables to evaluate the performances of our single-atom detection algorithm (section 2.3). Fig.(12c) presents the result of this computation applied to 10000 simulated FIM images produced every 50 evaporated atoms. Recalculated impacts positions are very close to the expected ones (Fig.(12a)). We may note that some additional depletions are produced at the centre of low index poles, and along major zone axes. Contrasts observed by simulation are very similar in intensity and location to contrasts experimentally produced (Fig.(12d)).

### 3.2 Detection efficiency comparison

Since the model provides *a priori* knowledge of the initial volume, we can measure the efficiency of the detection algorithm introduced above. Fig.(14a) shows ideal 3DFIM hitmap produced by the model. Fig 14b shows the same hitmap but calculated using the 3DFIM detection algorithm, using the sequence of 10000 produced FIM images during field evaporation. Fig14a is the image that should be obtained through a perfect detection of ion impacts. A slight degradation of the spatial resolution is observed, which is estimated to be in the range of half a pixel in the FIM images. The main visual artefact is the presence of depleted zones along the zones axes and the poles. To estimate the detection efficiency of the algorithm, in each pixel of the two maps, the number of hits was calculated, and the ratio of both numbers is calculated and the resulting two-dimensional map of this ratio is displayed in Fig.(14c). The detection efficiency approach 100% across most of the image. Along the zone lines and at pole centres, the performances of the algorithm are degraded. More advanced processing of the images may improve this efficiency, since most of the loss relates to the incapacity of the algorithm to separate some correlated field evaporated events giving rise to simultaneous impacts. The model is here an essential tool to optimize our detection algorithm and to estimate the best analysing conditions in 3DFIM (evaporation rate, detector spatial CCD resolution, etc…).

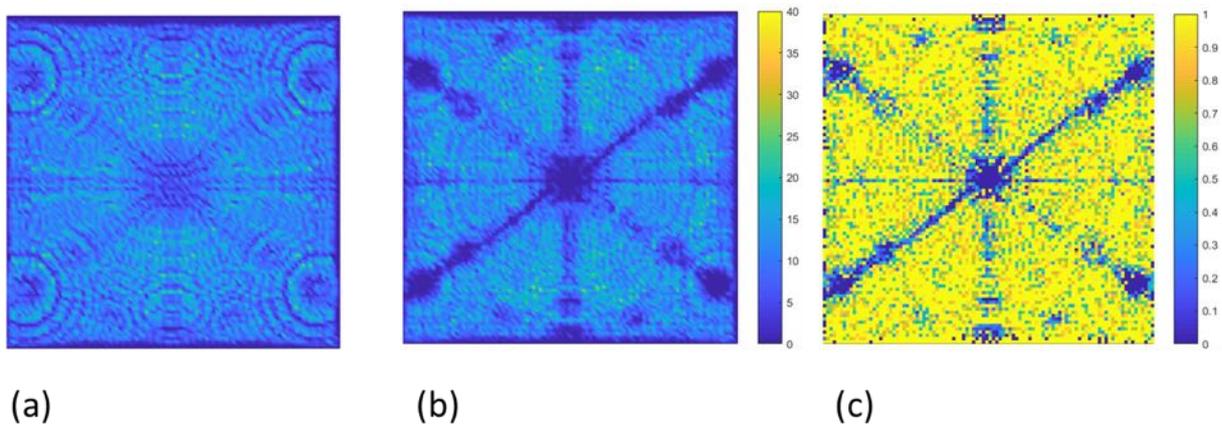

*Figure 14* : *Comparison of the local density of impacts from the ideal 3DFIM hitmap (a) and from 3DFIM detection algorithm (b) (colorscale indicates the number of hits per pixel ) (c) map of the detection efficiency of the algorithm calculated by dividing both images (yellow indicates 100% efficiency regions, dark blue 0% detection efficiency regions)*

3.3 Spatial resolution performances

The model developed in this paper can be used to estimate the spatial resolution loss introduced by the whole sequence of field evaporation, image formation, and algorithm detection. It was shown in section 2 that the experimental resolution could be in the range of a fraction of an Å in depth. The lateral resolution of the 3DFIM, is in some places sufficient to laterally separate individual atoms on a low index atomic planes. (011) Atomic planes are observed laterally both experimentally and by simulation, indicating a lateral resolution better than 0.2 nm (Fig 7). This resolution is better than the intrinsic resolution of FIM, limited by the size of atoms on the FIM screen. This performance is possible because 2 adjacent atoms laterally are never exactly at the same depth in the specimen. In order to map more accurately the spatial resolution performances, and to estimate the impact of the field imaging process on the spatial resolution the Fourier based method described in [1,46] was used on the simulated and experimental reconstructions. Using this method, a 3D map of the spatial resolution is obtained by measuring the amplitude of the diffraction peak corresponding to the (011) atomic planes aligned with the tip apex. The Fourier calculation is performed on all the atoms in a small sphere of 1nm in radius centred on each atom successively. Assuming a Gaussian dispersion of atoms around their correct position (atomic plane position), it was demonstrated that the Fourier amplitude $|F(k_{011})|$ is dependant of the spatial precision $\Delta_z$. Using $\Delta_z$ the full width half maximum of the spatial resolution in the real space, it was demonstrated that

$$\Delta_z \sim \frac{2d_{011}}{\pi}\sqrt{(\ln(2) \times \ln(|\mathbf{F}(\mathbf{k_{011}})|))} \qquad (11)$$

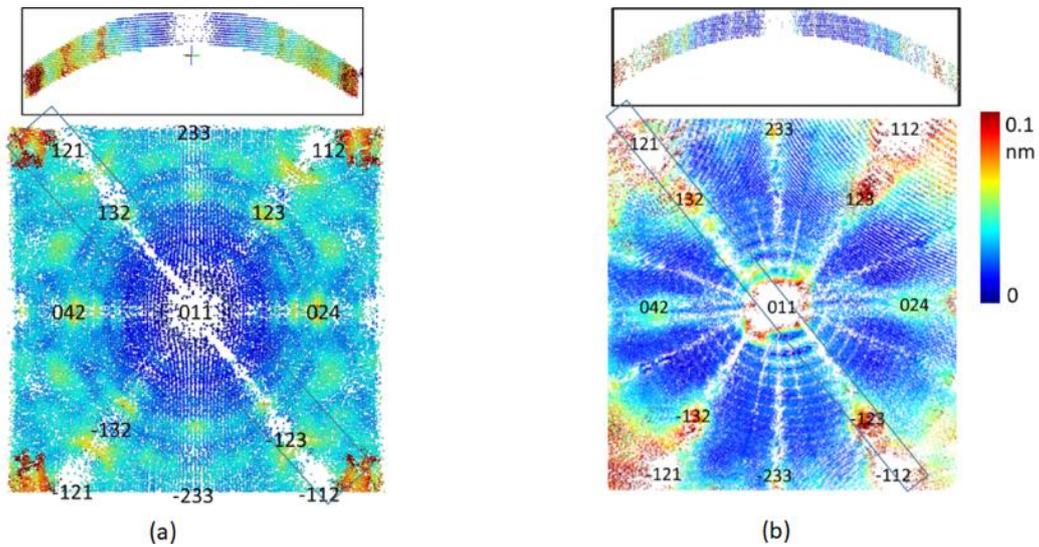

*Figure 15 : (a) Map of spatial resolution along [011] direction calculated on the 3DFIM tungsten simulation. Upper inset shows a slice in the reconstructed volume across the (011) pole. (b) Experimental map of spatial resolution along [011] direction calculated on the 3DFIM tungsten experiment (Blue spatial precision < 0.02 nm, Red thickness of atomic planes >0.1 nm). Upper inset shows a slice in the reconstructed volume across the (011) pole. . Measurement was made on $10^5$ reconstructed atoms*

Eq. 11 was used for experimental and simulated datasets (Fig.(15)) showing a similar behaviour. The best resolution is obtained when (011) atomic planes are parallel to the tip's surface with a minimum value of about 10pm in both experiment and simulation. A slight degradation of the resolution occurs with the distance to the centre of the pole. Indeed, the resolution along a given direction is the quadratic sum of the projection of the depth resolution and the projection of the lateral resolution along this particular direction, as observed experimentally in APT [46–48], albeit here with a much better lateral resolution for the FIM than the APT. We may note that (011) atomic planes are easily visible up to an angle of 40°. Some concentric variation of the resolution are also produced due to the ring distortion described previously. Some additional degradations of the resolution are observed close to pole centres, and in correlation with concentric rings artefacts. These degradations are probably related to the rapid evaporation of atoms on pole centres that may cause perturbation in the regular sequence of field evaporation.

## 4  Concluding remarks

In the present article, a general model for simulating the imaging process in the field ion microscope is proposed. Conversely to old geometrical model usually used to provide an interpretation of images, this model integrates both the dynamic process of field evaporation considering classical field evaporation theory, and the process of field ionization near the specimen's surface, to reproduce the sequence of FIM images currently observed experimentally. This model is used to reproduce the imaging process of the last generation of 3DFIM. The atom-by-atom, gradual evolution of FIM images is found to have a significant effect on the projection law in the FIM, even if ions are emitted a few angstroms from the surface, at the critical distance of field ionization. Zones lines artefacts and slight distortions of the images are highlighted. We however demonstrated that the spatial resolution of the instrument can achieve 10pm, and the detection efficiency can achieve 100% in defined zones of the tip surface. This ability validates the possibility to achieve observation of very small crystallographic defects such as vacancy clusters. Note that this ability was used as early as in the 60's and 70's to directly image the early stages of radiation damage, manifesting as very small, atomic-scale defects in the crystal lattice. With the help of the modelling approach future works to interpret contrasts from various atomic scale feature are now opened such as dislocations, dislocations loops, or small precipitates.  This approach can be also further enhanced to give correct interpretation of contrasts generated when analysing simple alloys (binary or ternary). It will help the development of a future 3DFIM with analytic performances.

**Acknowledgments**

The work was funded through EMC3 Labex-FEDER DYNAMITE and the EQUIPEX ANR-11-EQPX-0020 (GENESIS). We also acknowledge the French platform METSA. Simulated samples were generated using the ATOMSK software [49]. Visualization was performed using OVITO software [50]. BG and SSK are grateful for the financial support the Max-Planck Gesellschaft via the Laplace project. BG and FV acknowledges financial support from the ERC-CoG-SHINE-771602. FV thanks also the financial support of the University of Rouen through a CRCT funding.